%% file: main.tex
\documentclass[11pt]{article}
\usepackage{graphicx}
\usepackage[margin=1.25in]{geometry}
\usepackage[usenames,dvipsnames]{color}
\usepackage{url}
\usepackage[colorlinks = true,
            linkcolor = blue,
            urlcolor  = blue,
            citecolor = blue,
            anchorcolor = blue]{hyperref}

\newcommand\pubnumber{arXiv:2203.07501 [hep-ex]}
\newcommand\pubdate{\today}

\def\Title#1{\begin{center} {\LARGE #1 } \end{center}}
\def\Author#1{\begin{center}{ \sc #1} \end{center}}

\newcommand\pubblock{\rightline{\begin{tabular}{l} \pubnumber\\
         \pubdate \end{tabular}}}
\newenvironment{Abstract}{\begin{quotation} \begin{center}
                       ABSTRACT
     \end{center}\bigskip  }{\end{quotation}}

\input workshopsymbols.tex

\newcommand\snowmass{\begin{center}\rule[-0.2in]{\hsize}{0.01in}\\\rule{\hsize}{0.01in}\\
\vskip 0.1in Submitted to the  Proceedings of the US Community Study\\ 
on the Future of Particle Physics (Snowmass 2021)\\ 
\rule{\hsize}{0.01in}\\\rule[+0.2in]{\hsize}{0.01in} \end{center}}

\begin{document}

\pubblock

\Title{SoLAr: Solar Neutrinos in Liquid Argon}

\Author{Saba Parsa, Michele Weber, \it University of Bern, Switzerland}

\Author{Clara Cuesta, Inés Gil-Botella, Sergio Manthey, \it CIEMAT, Spain}

\Author{Andrzej M. Szelc, \it University of Edinburgh, United Kingdom}

\Author{Shirley Weishi Li, \it  Fermi National Accelerator Laboratory, Batavia, Illinois, USA}

\Author{Marco Pallavicini, \it Univ. of Genova and INFN Genova}

\Author{Justin Evans, Roxanne Guenette, David Marsden, Nicola McConkey, Anyssa Navrer-Agasson, Guilherme Ruiz, Stefan S\"oldner-Rembold~\footnote{stefan.soldner-rembold\@manchester.ac.uk}, \it University of Manchester, United Kingdom}

\Author{Esteban Cristaldo, Andrea Falcone, Maritza Delgado Gonzales, Claudio Gotti, Daniele Guffanti, Gianluigi Pessina, Francesco Terranova, Marta Torti, \it University of Milano-Bicocca and INFN, Italy}

\Author{Francesco Di Capua, Giuliana Fiorillo, \it University of Naples "Federico II" and INFN Napoli}

\Author{John F. Beacom, \it Ohio State University, Columbus, OHio, USA}

\Author{Francesco Capozzi, \it
Instituto de Fisica Corpuscular, Universidad de Valencia \& CSIC, Spain}



\newpage

\begin{Abstract}
\noindent SoLAr is a new concept for a liquid-argon neutrino detector technology to extend the sensitivities of these devices to the MeV energy range - expanding the physics reach of these next-generation detectors to include solar neutrinos. 
We propose this novel concept to significantly improve the precision on solar neutrino mixing parameters and to observe the ``hep branch'' of the proton-proton fusion chain. The SoLAr detector will achieve flavour-tagging of solar neutrinos in liquid argon.
The SoLAr technology will be based on the concept of monolithic light-charge pixel-based readout which addresses the main requirements for such a detector: a low energy threshold with excellent energy resolution ($\approx 7\%$) and background rejection through pulse-shape discrimination. 

The SoLAr concept is also timely as a possible technology choice for the DUNE ``Module of Opportunity'', which could serve as a next-generation multi-purpose observatory for neutrinos from the MeV to the GeV range. 
The goal of SoLAr is to observe solar neutrinos in a $10$ ton-scale detector and to demonstrate that the required background suppression and energy resolution can be achieved. SoLAr will pave the way for a precise measurement of the $^8$B flux, an improved precision on solar neutrino mixing parameters, and ultimately lead to the first observation of hep neutrinos in the DUNE Module of Opportunity.
\end{Abstract}


\snowmass

\newpage

\def\thefootnote{\fnsymbol{footnote}}
\setcounter{footnote}{0}
\section{Introduction}
The next decade will be a turning point in neutrino physics as facilities with detectors of unprecedented mass and sensitivity become operational. These deep-underground detectors will observe neutrinos from both ``artificial" and natural sources. Artificial neutrinos from accelerators are currently the main drivers for the construction of underground experiments since they can underpin theories that explain the mechanism of creating the matter-over-antimatter imbalance in our universe. 

The need to observe neutrinos in a deep-underground location to mitigate backgrounds also opens up a wealth of new science opportunities that have not been exploited to date due to current limitations in technology and physics understanding. 

A strong case for extending the physics reach of future liquid-argon detectors to low-energy physics has been made in the White Paper of the LEPLAr team~\cite{leplar}.

We aim to exploit this particularly exciting science opportunity using the novel SoLAr concept. Our goal is to observe the last and most elusive process that creates neutrinos in the Sun, the fusion of protons and helium nuclei 
$$^3 {\rm He} + p \to  ^4 {\rm He} + e^{+} + \nu_e ,$$ 
the ``hep branch" of the proton-proton fusion chain. The neutrino flux from the hep branch is so faint that no experiment has so far been able to observe it. An upper limit on the integral total flux of hep neutrinos was established by the SNO collaboration~\cite{snohep}.
The first observation of hep neutrinos would have a tremendous impact on astro-particle physics, specifically on our understanding of stellar evolution and the physics of massive neutrinos, since this process is driven by a much larger fusion core than any other reaction producing neutrinos in the Sun. 

A facility capable of observing such a tiny flux ($7.98 \pm 2.39 \times 10^3$ cm$^{-2}$s$^{-1}$ according to the B16-GS98 solar model~\cite{1}) could also provide the first high-precision measurement of the $^8$B neutrino flux decoupled from the uncertainties coming from neutrino oscillations. We propose a novel integrated charge-light readout concept for liquid-argon detectors to perform such measurements. The precise comparison of the neutrino-mixing parameters measured with SoLAr versus reactor experiments will open up an entirely novel opportunity to test new physics models.

Furthermore, neutrino emission is the main energy-loss mechanism during supernova explosions, ejecting neutrinos with energies similar to those of hep neutrinos. The first hep neutrino observatory is thus by construction also the most precise experiment to detect supernova explosions. Its data can be combined with visible, X-ray, and gamma-ray, and potentially also gravitational-wave information, to provide a complete picture of the supernova collapse, one of the least understood processes in the visible universe.       

Discoveries and precision in the observation of celestial bodies always follow from a major technology breakthrough. The observations of supernova neutrinos and hep neutrinos are no exception. To trace the stellar evolution in the aftermath of a core-collapse, we need to identify the neutrinos that are produced by the exploding star, measure their flavour, and identify the flavour at the source. Extracting this information is further complicated by neutrino oscillations that occur when neutrinos traverse the dense matter of the exploding star. 

Solar neutrinos are also affected by MSW oscillations~\cite{msw1, msw2}
in the core of the Sun. We thus need a breakthrough to construct an experiment capable of measuring neutrinos in the energy range $1$--$20$~MeV, between the high-energy hep Q-value and the low-energy $^7$Be line, the region of the onset of "MSW resonance"
in the Sun.
Achieving this breakthrough is the goal of the SoLAr concept.

\section{Flavour tagging for solar neutrinos}
Following the original solar-neutrino experiment by R.~Davis et al. in the Homestake mine, the Kamiokande collaboration started to observe solar neutrinos in a large-mass detector using purified water. Solar neutrinos with energies of a few MeV scatter off water electrons producing visible Cherenkov rings recorded by large area photomultipliers. The technique was pioneered by M. Koshiba in the 1980s and led to the observation of neutrinos produced by the $^8$B decay in the Sun and by the explosion of a Type II supernova in 1987 (SN1987A)~\cite{2,3}, a discovery that was recognized by the 2002 Nobel Prize in Physics. 

The observation of elastic scattering of neutrinos on electrons in water cannot discriminate between neutrino flavours since the cross section is nearly flavour blind. Flavour tagging can, however, be achieved by combining two detection processes. The neutral-current process is flavour blind. It therefore provides the total flux of neutrinos from the celestial object, while the flavour-sensitive charged-current process tags the subsample of a specific flavour, e.g., electron (anti)neutrinos. This technique provided definitive evidence of solar neutrino oscillation but, to date, has been implemented only in moderate mass detectors like SNO~\cite{4}. 

Flavour tagging can be scaled to masses even bigger than the Kamiokande detector replacing water with liquid argon (LAr). This technique was proposed long ago by J.N. Bahcall et al.~\cite{5} and in Ref.~\cite{6}. It exploits the flavour-blind scattering of neutrinos off the electrons of argon atoms ($\nu + e^- \to
\nu + e^-$) and the charged-current electron-neutrino interaction with $^{40}$Ar 
($\nu_e + ^{40}{\rm Ar} \to e^- + ^{40}{\rm K}^{*}$). 

\section{A Module of Opportunity}
In the forthcoming decade, the DUNE modules constructed as liquid-argon Time Projection Chambers (LArTPCs) at SURF in South Dakota will be scaled to masses of 10,000 tons and more~\cite{7}. The long-baseline neutrino facility (LBNF) at SURF will comprise two main caverns with space for up to four such DUNE modules. While the first two modules are currently under construction and will be operational before the end of this decade, no technology has yet been chosen for the third and particularly for the fourth module. The latter is therefore denoted as the ``Module of Opportunity", as the goal is to extend the physics reach of the experiment with novel technologies, while still retaining the sensitivity to beam neutrino interactions needed to reach the flagship DUNE physics goals of discovering leptonic CP violation and testing the three-flavour paradigm. 

The design of the first DUNE modules can not achieve the sensitivity needed for a first observation of hep neutrinos.
The authors of Ref.~\cite{8} point out that DUNE, has the potential to deliver world-leading results for  solar neutrinos, with a realistic hope of separately measuring the flux, including the hep neutrino flux, and mixing parameters for reasonable choices of detector properties and contemporary mixing parameters. However, this ambitious goal also requires “breakthroughs in detection strategy.” 

The goal of the SoLAr proposal is to deliver such breakthroughs and to experimentally prove that the technique of Ref.~\cite{6} can be ported to a large-mass experiment, the future DUNE Module of Opportunity~\cite{9}. This development comes at a perfect time to leverage recent promising advances in pixel and light readout technology for LArTPCs. The timing is also perfectly aligned with the DUNE Module of Opportunity process. 

The Module of Opportunity will have to be a multi-purpose LAr detector covering the full spectrum of neutrino interactions so that it can serve as an international neutrino observatory for a broad and growing science community. To achieve the required technology breakthrough that will allow us to detect solar neutrinos, we need to address the limitations of current LAr detectors and demonstrate the feasibility of the SoLAr concept in an underground laboratory.

\section{Liquid-argon detectors for MeV neutrino detection}

Liquid argon offers simultaneously a target medium for neutrino scattering, high-precision reconstruction of particles produced after the scattering by collecting the ionization electrons~\cite{10}, and a remarkable amount of vacuum ultraviolet (VUV) light generated by scintillation ($\approx 25,000$--$40,000$ photons/MeV at $128$ nm wavelength depending on the electric field configuration)~\cite{7}. 

A target mass of $\approx 10$~kt offers a superb yield of solar neutrino events with about 10,000 $\nu_e$ charged-current interactions per year with energies above $5$ MeV that can be uniquely attributed to the $^8$B reaction, and about 2,000 $\nu_e$ scattering events per year used to decouple oscillation effects and reach the best measurement ever of the $^8$B flux. An ultra-pure sample of hep neutrino interactions above 14 MeV (up to 50 events per year) would allow us to observe the faintest solar neutrino source, the proton-helium fusion chain, for the first time. We have therefore identified a set of requirements that are critical to achieve such sensitivity with a solar-neutrino liquid-argon detector:
\begin{itemize}

\item Effective suppression of radioactive background from the detector material and the surrounding rock, since these backgrounds, if unmitigated, exceed the faint solar neutrino signal by orders of magnitude. At energies $>5$~MeV, the background is dominated by neutrons that produce photons through de-excitation of argon, whereas the $^{40}$Ar($\alpha,\gamma$) process that creates photons by the scattering of $\alpha$ particles from radon decays becomes critical at higher energies (around $15$~MeV)~\cite{11}. 

\item Efficient reconstruction for low-energy neutrino interactions, since the hep neutrino flux lies in the energy range between about 5 and 18.8 MeV, and reconstruction of directionality, since solar neutrinos are produced from a “point source” at a distance of $10^8$ km from the detector, and directionality can potentially help with signal identification. 

\item Effective discrimination between the components of the solar neutrino flux and identification of hep neutrinos, since most of the $^8$B neutrino flux lies in the $5$--$10$~MeV energy region, with a cut-off at $17$~MeV. This is just below the hep-neutrino cutoff of $18.8$~MeV, and opens a small $1.8$~MeV window to observe a pure sample of hep neutrinos~\cite{12}. Excellent energy resolution is critical to achieve this separation. This will require combining information from charge and light~\cite{LArIAT} as well as tagging of blips resulting from nuclear de-excitation gammas and neutrons~\cite{Gardiner},\cite{LittleJohn}, which has been demonstrated experimentally~\cite{ArgoNeuT}.
The current generation of massive LAr detectors is not equipped to meet these requirements.

\item Fiducialization and self-shielding is achieved in large detectors by selecting a fiducial volume located in the inner part of the liquid argon chamber and using the outer LAr shell to veto events originating from the surrounding. Modern LAr detectors use commercial cryostats originally built for the transportation of cryogenic liquids (“membrane cryostats”~\cite{7}), which are neither radiopure nor effective in shielding neutrons from the underground experimental area. The current membrane cryostats employed at CERN for the validation of DUNE (NP04/ProtoDUNE~\cite{13}) require a tight fiducialization, which suppresses the hep neutrino yield by a factor of 10 and makes the first detection of hep neutrinos impractical with the current DUNE module design. 
\item``Pulse shape discrimination (PSD)" is the workhorse for background suppression in direct dark matter detection in LArTPCs~\cite{14}. PSD exploits the time distribution of light to discriminate between electrons and more ionizing, heavier, particles (such as $\alpha$ particles). Unlike dark matter detectors, the detected light signal in a kton-scale LArTPC is too weak for efficient PSD, unless we add, for example, reflective foils coated with wavelength-shifting films. To further enhance light output, xenon doping of the argon is planned for the first two DUNE modules. Doping changes the time structure of the light signal and makes it impossible to exploit PSD for background discrimination.
\item	Minimum number of hits requirements in wire-based detectors introduce track length requirements and a corresponding lower cut-off of about 10 MeV for tracks, while the limited photon detector coverage limits the energy resolution. 
\end{itemize}

\begin{figure}
\begin{center}
\includegraphics[width=0.80\hsize]{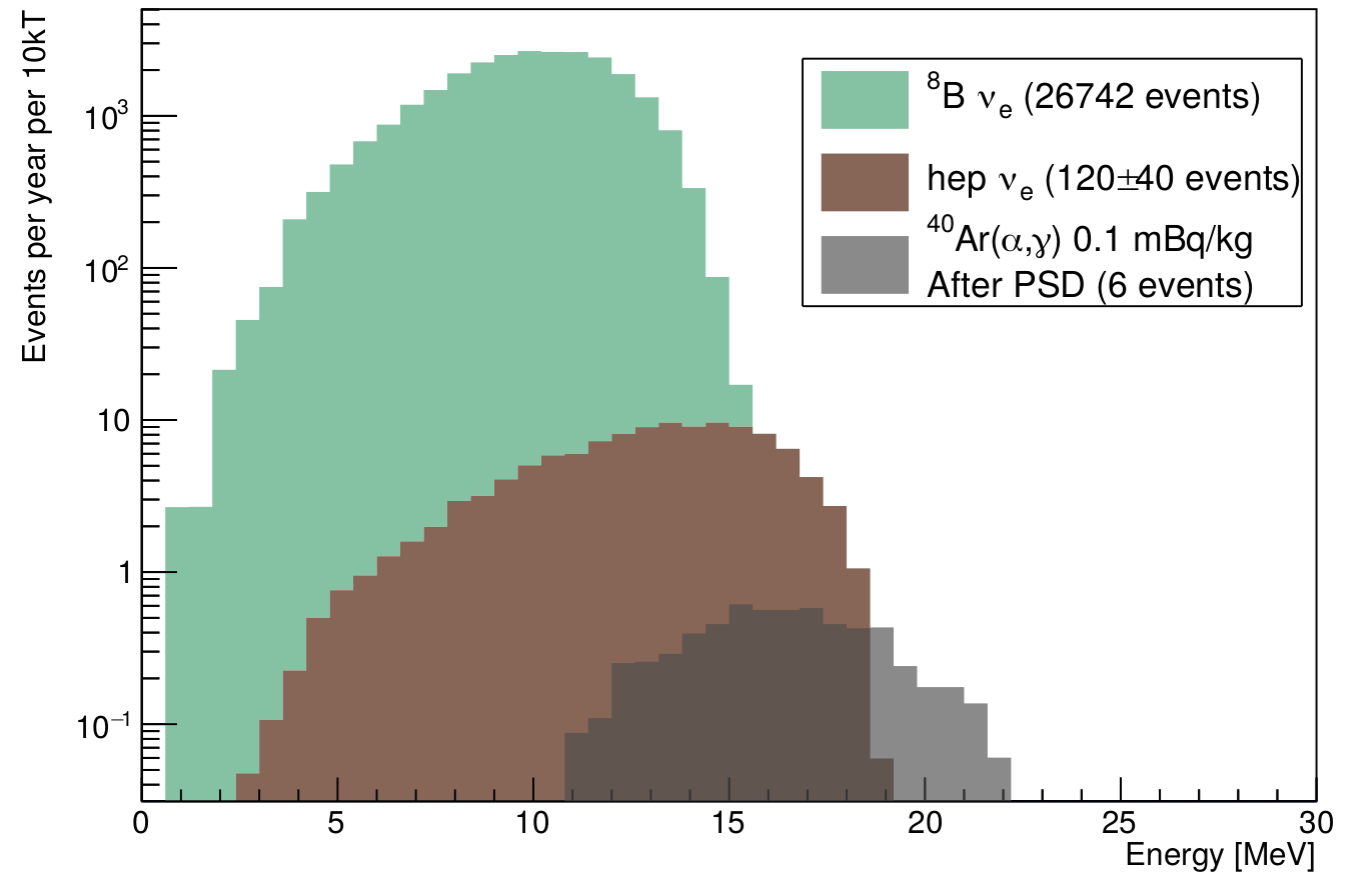}
\end{center}
\caption{Solar-neutrino energy spectrum in a LArTPC for $^8$B and hep neutrinos with radon-induced background after applying pulse shape 
discrimination.}
\label{fig:fig1}
\end{figure}

The authors of Ref.~\cite{15} have studied the remarkable impact of energy resolution in the background budget of a LArTPC with conventional membrane cryostats. The combination of shielding and a $7\%$ energy resolution used in Ref.~\cite{15} gives access to the $5$--$10$~MeV region, where most of the $^8$B neutrinos reside. The energy resolution is instrumental to sharpen the $17$~MeV cutoff of the $^8$B neutrino spectrum, as shown in Figure~\ref{fig:fig1}, which lies just below the hep cutoff of $18.8$~MeV, and opens a $1.8$~MeV window to observe a pure sample of hep neutrinos [12]. In Figure~\ref{fig:fig1}, we also show the expected level of $^{40}$Ar($\alpha,\gamma$) background in the higher-energy region after applying a pulse shape discrimination cut. The pulse shape discrimination was performed using the light simulation described in Ref.~\cite{diego}, with an optical timing resolution of $100$~ns. The signal spectra assume perfect detector efficiency and energy resolution. 

\section{The current state-of-the art DUNE modules}

The conventional charge collection system of a LArTPC relies on the original design proposed by C. Rubbia in 1977 as demonstrated by ICARUS~\cite{16} and other operating LArTPCs (e.g., MicroBooNE~\cite{uB}, ProtoDUNE~\cite{13}). The system is based on thousands of several metre long Cu-Be wires arranged in three planes oriented at different angles. The wires are held on large modular structures called Anode Plane Assemblies (APAs)~\cite{7}, which are currently under construction for the first DUNE module in the US and the UK. 
The APA pitch of typically 3-5 mm, i.e, the distance between wires within one of the planes, is constrained by the diffusion of the drifting charge. The detector provides 2D images which need to be combined for 3D reconstruction. The photon detection system (PDS) is located behind the wires and covers just a small fraction ($12\%$) of the APA area. 

For the second DUNE module, printed circuit boards (PCBs) with strips will replace the wires, which simplifies construction and installation with similar performance expected as for the wire-based detector. The PDS system will be installed on the sides of the cryostat and possibly on the cathode. Both modules will use the X-ARAPUCA~\cite{18} system for light detection. The argon will be doped with xenon to increase light yield.

\section{The SoLAr concept of integrated charge-light readout}

The SoLAr concept is capable of simultaneously addressing these challenges and meeting the physics requirements by a novel design of the charge collection and photon detection systems of the LArTPC. This novel LArTPC concept is expected to achieve a precision on the measurement of the $^8$B flux of $2.5\%$. With a detector of the size of the DUNE Module of Opportunity, it reaches sensitivity for $5\sigma$ evidence of the occurrence of the hep process, measures the corresponding neutrino flux at the $10\%$ level, and successfully addresses the claim of Ref.~\cite{8}: “no other experiment, even proposed, has been shown capable of fully realizing these discovery opportunities”.

The traditional wire planes used for LArTPCs are replaced by a modular pixel design, whose backbone implements the same multiplexing scheme used in modern high-density cameras. The possibility of replacing the wire-based APAs with a pixelated system in a modular design has already been demonstrated for the DUNE Near Detector. The smaller Near Detector is located on the surface at Fermilab, close to the neutrino source, and thus has high occupancy. The charge collection is embedded in a PCB, directly coupled to the front-end electronics. The system has been adopted by DUNE as the charge readout system of the liquid-argon near detector (NDLAr)~\cite{17}.  

In parallel, institutions from the DUNE Photon Detection System Consortium have developed a new class of materials and dyes to enhance the performance of the PDS.
In 2020, a joint research programme between liquid-argon detector scientists and one of our industrial partners (Hamamatsu Photonics) delivered a Silicon Photomultiplier (SiPM) that reached a record efficiency ($30\%$ PDE) for $128$~nm light at the argon boiling point (87K). Nearly at the same time, the first integrated system for multiplexing the SiPM signal was commissioned and operated inside strong electric fields. 

The SoLAr readout unit will be a pixel tile in CMOS technology that embeds charge readout pads located at the focal point of the LArTPC field shaping system and collects VUV photons in thousands of microcells operated in Geiger mode. Each monolithic sensor produces analog signals corresponding to the charge of the electrons collected in each pixel and to the photons collected in the pixel frame, respectively. 

It is not possible to deliver such an ambitious design for an integrated charge-light solar neutrino telescope relying on conventional particle detectors. To accomplish the metal-to-semiconductor transition in large mass neutrino detectors, we must first master photonics, then integrated microelectronics, and finally astrophysics. These needs drove the creation of the SoLAr Collaboration.

\section{The SoLAr Detector}

\begin{figure}
\begin{center}
\includegraphics[width=0.80\hsize]{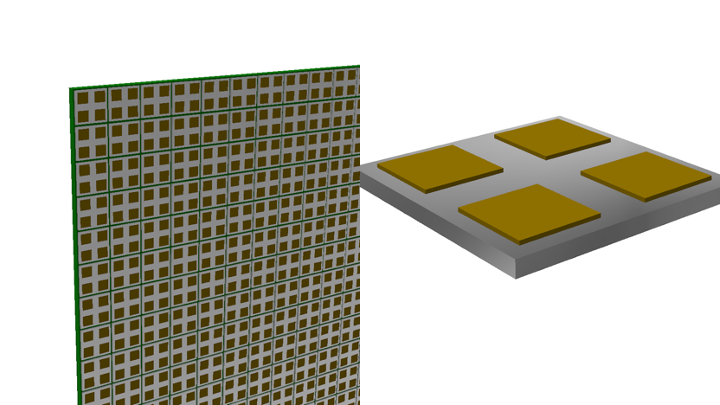}
\end{center}
\caption{
 (right) The all-silicon readout unit is based on VUV SiPMs with charge collection pads on the same unit. The unit corresponds to one MPPC photon detection readout channel and four charge readout channels. (left) a SoLAr tile of 32 cm x 32 cm with $50 \times 50$ full silicon units.
}
\label{fig:fig2}
\end{figure}

The SoLAr Collaboration proposes to build the SoLAr detector, a neutrino telescope at the 10 ton scale where charge and light are recorded by an integrated all-silicon anode plane. At the same time, the SoLAr detector will be a crucial step towards proposing a full-scale detector. We propose to install the SoLAr detector in the Boulby Underground Laboratory in the UK to observe --- for the first time --- flavour-tagged solar neutrinos in liquid argon. SoLAr will not only demonstrate the concept of the monolithic light-charge readout but addresses all challenges mentioned above. 
The all-silicon readout unit is based on VUV SiPMs with charge collection pads on the same unit. The unit corresponds to one MPPC photon detection readout channel and four charge readout channels. They are then assembled to SoLAr tiles of $32\,\mathrm{cm} \times 32\,\mathrm{cm}$ with $50\times 50$ full silicon units. This is the fundamental detector element that will be tested as prototypes and then integrated in readout planes for SoLAr as shown in Figure~\ref{fig:fig2}. 

We aim for an active volume of $1.6\,\mathrm{m}\times 2\,\mathrm{m}\times 2.6\,\mathrm{m}$ corresponding to an active mass of liquid argon of 11.5 tons. SoLAr is imagined to be a horizontal drift LArTPC whose geometry resembles the design of the first module of DUNE. The internal design, however, is completely novel: the wire planes are replaced by a matrix of pixels combined with a light collection efficiency four times larger than for an instrumented wire-based APA. Unlike conventional LArTPCs, light is collected over five of the six surfaces, with photon detectors also installed behind the field cage (lateral light detectors). 
The field cage of SoLAr employs the novel design of the second DUNE module (“vertical drift”) and benefits from a $70\%$ transparency to scintillation light. Additional light detectors will be added behind the field cage based on the X-ARAPUCA concept~\cite{18}, but with twice the efficiency of existing devices. The new dichroic filters for the X-ARAPUCA will be used for enhanced light detectors. 

The cryostat will be designed to solve the long-standing issue of poor neutron moderation in LAr while retaining the compactness of the detector. The SoLAr detector relies on a membrane cryostat similar to the conventional modules of DUNE. However, we propose a novel design that abates the need for fiducialization because thermal insulation is achieved by radiopure materials, and neutron absorption is accomplished by low radioactivity hydrogenated layers located at both ends of the membrane. The material selection is based on radiopurity, cryo-resilience, and thermal conductivity.

SoLAr will require exquisite material screening. The monolithic pixels are intrinsically pure due to the all-silicon design and the use of radio-pure silicon photomultipliers (SiPMs). The most critical components are the cryostat walls, the field cage and mechanical supports. Radiopurity screening will be performed at the collaborating institutions.


\section{The science goals of the SoLAr concept}

The SoLAr concept is based on a fully integrated system that records the argon scintillation light to perform pulse shape discrimination and simultaneously the ionization charge that provides a superior energy resolution ($7\%$ at few-MeV) and mm-scale ``true" 3D reconstruction. It is designed to address the main requirements for an ambitious physics programme using low-energy neutrinos: background rejection, high efficiency in the signal energy region, and excellent energy resolution for identification of the solar flux components. SoLAr aims to achieve in LArTPC a technology leap that mirrors the most important breakthrough of high energy physics in the 1990s: the transition from wire-based particle trackers to pixelated all-silicon devices.  

The initial science goals of SoLAr are to assess self-shielding capabilities including virtual fiducialization and to assess the expected performance of the SoLAr detector (background reduction with mitigation algorithms and pulse-shape discrimination to a background-free experiment above 14 MeV, and $7\%$ energy resolution). 

A prototype will be needed to validate the performance using the SoLAr detector and observe solar neutrinos ($^8$B) with SoLAr at $>5\sigma$ significance. The prototype will also enable us to estimate the sensitivity to solar neutrinos (target: $2.5\%$ measurement of $^8$B and $11\%$; measurement of hep flux) for the Module of Opportunity to demonstrate the superior detection capability for low-energy neutrinos of this technology.

\end{document}

%% file: workshopsymbols.tex
\def\beq{\begin{equation}}
\def\eeq#1{\label{#1}\end{equation}}
\def\eeqn{\end{equation}}

\newenvironment{Eqnarray}%
   {\arraycolsep 0.14em\begin{eqnarray}}{\end{eqnarray}}
\def\beqa{\begin{Eqnarray}}
\def\eeqa#1{\label{#1}\end{Eqnarray}}
\def\eeqan{\end{Eqnarray}}

\let\bar=\overbar

\def\lsim{\mathrel{\raise.3ex\hbox{$<$\kern-.75em\lower1ex\hbox{$\sim$}}}}
\def\gsim{\mathrel{\raise.3ex\hbox{$>$\kern-.75em\lower1ex\hbox{$\sim$}}}}

\def\del{\partial}
\def\Dslash{\not{\hbox{\kern-4pt $D$}}}
\def\dslash{\not{\hbox{\kern-2pt $\del$}}}
\def\pslash{\not{\hbox{\kern-2pt $p$}}}
\def\ETmiss{\not{\hbox{\kern-4pt $E$}}_T}

\def\Dlr{\mathrel{\raise1.5ex\hbox{$\leftrightarrow$\kern-1em\lower1.5ex\hbox{$D$}}}}

\def\MSB{{\bar{M \kern -2pt S}}}
\def\msb{{\bar{\scriptsize M \kern -1pt S}}}

\def\drb{{\bar{\scriptsize D \kern -1pt R}}}

